\documentclass[
aps,
prl,
reprint,
superscriptaddress,
longbibliography,
nofootinbib
]{revtex4-2}

\usepackage{amsmath,amssymb,mathtools,bm}
\usepackage{microtype}
\usepackage{xurl}
\usepackage{tikz}
\usetikzlibrary{positioning}

\newcommand{\T}{\mathbb T}
\newcommand{\R}{\mathbb R}
\newcommand{\Z}{\mathbb Z}

\newcommand{\cH}{\mathcal H}
\newcommand{\cM}{\mathcal M}

\newcommand{\MCG}{\mathrm{MCG}}
\newcommand{\PU}{\mathrm{PU}}
\newcommand{\Kah}{\mathrm{K\ddot{a}hler}}
\newcommand{\real}{\mathrm{real}}

\newcommand{\wt}[1]{\widetilde{#1}}

\usepackage{xcolor}
\definecolor{linkblue}{RGB}{65,125,205}

\usepackage[
    colorlinks=true,
    linkcolor=linkblue,
    citecolor=linkblue,
    urlcolor=black,
]{hyperref}

\begin{document}

\title{Torus Berry Data Determine All-Genus Abelian Topological Orders}
\author{Daniel Galviz}
\affiliation{Yau Mathematical Sciences Center, Tsinghua University, Beijing, China}

\begin{abstract}
We show that for Abelian Chern–Simons topological orders, torus Berry matrices determine the all-genus extended TQFT. We identify the topological part of the projective Berry holonomy under metric deformations with the mapping-class-group representation of the Abelian Chern--Simons TQFT and prove that the normalized torus data reconstruct its finite quadratic module $(G,q)$. Recent work showed that $(G,q)$ classifies the extended theory up to symmetric monoidal natural isomorphism, and that the associated Abelian defect extension is determined by the pointed modular category $\mathcal C(G,q)$ arising from the same finite quadratic-module data. Therefore genus-one Berry data determine not only the all-genus bulk theory but also its associated defect structures without choosing a $K$-matrix presentation. We also prove that, for normalized character row errors $\delta<21.96\%$, nearest-row decoding recovers the Abelian fusion algebra independently of the number of anyons. The result applies to Abelian fractional quantum Hall and spin-liquid phases described by even-lattice Chern--Simons theories.
\end{abstract}

\maketitle

\textit{\textbf{Introduction--}}
A central challenge in identifying a gapped topological phase is determining how much information is contained in its protected ground-state manifold. Unlike conventional phases, topological phases are characterized by Berry transport, braiding statistics, and geometric response rather than local order parameters. Wen and Keski-Vakkuri proposed that the Berry evolution of the degenerate ground-state manifold under changes of spatial geometry should encode the universal topological information of the phase~\cite{wen1990,KeskiVakkuriWen1993}. A fundamental question is whether these torus Berry data are sufficient to reconstruct the complete all-genus topological theory ~\cite{wen1990,KeskiVakkuriWen1993,wen2012,Wen2016}.

On a torus, the holonomies are the modular matrices $S$ and $T$, which can also be extracted from microscopic wave functions and tensor-network calculations~\cite{Liu2013,He2014,You2015,Hu2023}. For general non--Abelian order, genus one is not complete: inequivalent modular categories can share identical modular data~\cite{MignardSchauenburg}, and higher-genus or punctured-surface representations can distinguish such examples~\cite{wenwen2019}. The Abelian case is special because this question can be answered completely.

Although Abelian topological orders admit an algebraic classification by finite quadratic modules~\cite{Joyal-Street,DGNO2010}, it has remained unclear how this classification emerges directly from geometric Berry measurements. On the other hand, Abelian Chern--Simons theory, theta functions, and finite Weil representations provide equivalent realizations~\cite{BelovMoore,Stirling,Andersen,GelcaUribe,Galviz4}.  

In practical settings, the torus Berry matrices are precisely the quantities accessible from microscopic calculations, including tensor-network approaches and exact diagonalization of finite systems. Such modular data have also been reconstructed experimentally in quantum simulators, demonstrating their robustness as probes of topological order \cite{Luo2018}.

Recent work constructed the all-genus Abelian Chern--Simons theory as an extended\footnote{Here \textit{extended} refers to the Turaev--Walker bordism formalism for 2--3 TQFT \cite{Walker, Turaev1994}, in which boundary surfaces carry additional Lagrangian data and gluing is corrected by Maslov--Kashiwara indices.} TQFT \cite{Galviz2},  and subsequently proved its equivalence with Abelian Reshetikhin--Turaev theory and showed that the resulting extended theories are classified, up to symmetric monoidal natural isomorphism, by finite quadratic modules $(G_K,q_K)$ \cite{Galviz3,Galviz4}. This supplies the classification theorem needed for the Abelian $(2+1)$-dimensional case of the Wen–Keski-Vakkuri conjecture. Moreover, the finite quadratic module is not only the invariant controlling the bulk extended theory, but also the organizing invariant of its defect extension. Recent constructions of Abelian Chern--Simons defects identify boundaries, domain walls, and condensation sectors as structures associated with the pointed modular category $\mathcal C(G,q)$ and hence with the reconstructed quadratic module. Therefore, once the torus Berry data reconstruct $(G,q)$, they also determine the associated Abelian defect theory \cite{Galviz4,Galviz6}.

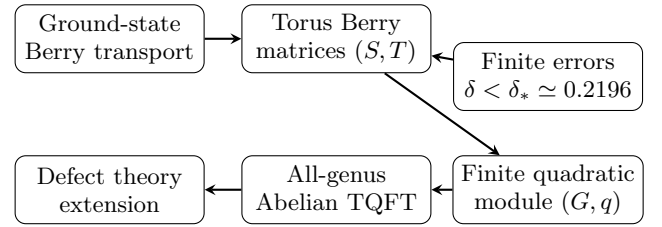
\begin{figure}[h!]
\centering

\begin{tikzpicture}[
    box/.style={
        rectangle,
        rounded corners,
        draw,
        align=center,
        minimum width=2.5cm,
        minimum height=0.9cm,
        font=\small
    },
    arrow/.style={
        ->,
        thick,
        >=stealth
    }
]

% Nodes
\node[box] (berry) at (-3,2)
{Ground-state\\Berry transport};

\node[box] (st) at (0,2)
{Torus Berry\\matrices $(S,T)$};

\node[box] (quad) at (2.8,0)
{Finite quadratic\\module $(G,q)$};

\node[box] (tqft) at (0,0)
{All-genus\\Abelian TQFT};

\node[box] (defect) at (-3,0)
{Defect theory\\extension};

% Main arrows
\draw[arrow] (berry) -- (st);
\draw[arrow] (st) -- (quad);
\draw[arrow] (quad) -- (tqft);
\draw[arrow] (tqft) -- (defect);

% Error box
\node[box] (noise) at (2.8,1.5)
{Finite errors\\
$\delta<\delta_*\simeq0.2196$};

\draw[arrow] (noise) -- (st);

\end{tikzpicture}

\caption{\footnotesize\textbf{Berry tomography map.}
Torus Berry transport reconstructs the finite quadratic module $(G,q)$ controlling the Abelian topological theory. The reconstructed quadratic module determines fusion and braiding and, by the classification of extended Abelian Chern--Simons theories, fixes the all-genus theory up to symmetric monoidal natural isomorphism. The same quadratic-module data determine the associated defect extension, including boundaries, domain walls, and condensation sectors. The reconstruction is stable under normalized-row errors through nearest-row decoding with universal threshold $\delta\simeq21.96\%$.}
\label{fig:berry_tomography}

\end{figure}

We establish a Berry tomography map in Fig.\ref{fig:berry_tomography} by identifying the projective Berry holonomy generated by metric deformations with the (real-polarization) extended TQFT representation. We also prove that exact torus modular matrices provide an intrinsic certificate for reconstructing the underlying anyon theory. We show that this reconstruction is stable against finite errors. Altogether, it provides an algorithmic procedure that extracts the underlying Abelian anyon theory directly from numerically computed Berry data, without requiring prior knowledge of the microscopic Hamiltonian or a chosen $K$-matrix description.

\vspace{5pt}
\textit{\textbf{Metric Berry bundle and the real-polarized TQFT--}}
Let $\T=\mathfrak t/\Lambda\cong U(1)^N$ and let $K:\Lambda\times\Lambda\to\Z$ be even, integral, symmetric, and nondegenerate. For a closed oriented surface $\Sigma_g$, the phase space of flat $\T$ connections is the integral symplectic torus
\begin{equation}
\begin{aligned}
\cM_\Sigma(\T)&=H^1(\Sigma;\mathfrak t)/H^1(\Sigma;\Lambda),\\
\omega_K([\alpha],[\beta])&=\int_\Sigma K(\alpha\wedge\beta).
\end{aligned}
\label{eq:phase}
\end{equation}
Choose a $K$-self-adjoint involution $\epsilon_K$ such that $g_K(u,v):=K(u,\epsilon_Kv)$ is positive definite. If $h$ is a metric on $\Sigma$, the Hodge star $j_h=*_{h}$ on $H^1(\Sigma;\R)$ satisfies $j_h^2=-1$, and
\begin{equation}
J_h=j_h\otimes\epsilon_K
\label{eq:J}
\end{equation}
defines a compatible positive complex structure on \eqref{eq:phase}. Half-form K\"ahler quantization therefore gives a finite-dimensional bundle $\cH^{\Kah}_{K,g}\to\mathcal T_g$ over Teichm\"uller space. Independently, a rational Lagrangian $L\subset H^1(\Sigma;\R)$ defines the real-polarized Bohr--Sommerfeld space $\cH^{\real}_{K,L}(\Sigma)$ used by the extended TQFT~\cite{Galviz2}.

For invariant nonnegative polarizations of an integral symplectic torus, half-form BKS maps are unitary and transitive~\cite{Baiermouraonunes}. Real-polarization quantization with half-densities instead carries a projective ambiguity measured by the Maslov--Kashiwara index; choosing metaplectic data lifts this projective construction to an ordinary Hilbert space~\cite{Manoliu1,Galviz2}. These are two realizations of the same projective quantization.  It therefore identifies geometric Berry transport with the intrinsic mapping-class-group representation of the Abelian TQFT.

\textbf{Theorem 1.}
Fix a metaplectic resolution of the canonical projective real-polarized quantization. For every $h\in\mathcal T_g$ and for every rational integral Lagrangian polarization $L$, half-form BKS transport induces a projective unitary isomorphism
\begin{equation}
B_{L,h}:\mathbb P\cH^{\Kah}_{K,h}(\Sigma_g)\longrightarrow
\mathbb P\cH^{\real}_{K,L}(\Sigma_g),
\label{eq:BKSproj}
\end{equation}
independent of the phase choice in the lift. Let $\rho^{\Kah}_{K,g}$ be the projective mapping-class-group representation obtained from the K\"ahler/BKS bundle, and let $\rho^{\real}_{K,g}$ be the projective representation obtained from the half-density real-polarized theory. Then
\begin{equation}
B_{L,h}\,\rho^{\Kah}_{K,g}(f)
=\rho^{\real}_{K,g}(f)\,B_{L,h}
\qquad\text{in }\PU(\cH)
\label{eq:intertwine}
\end{equation}
for every $f\in\MCG(\Sigma_g)$.

The proof is provided at the end. Its key point is that strict transitivity belongs to the half-form BKS system, whereas the Maslov–Kashiwara  phase appears when the same transport is written in the half-density real-polarization convention. The two descriptions have the same projectivization. 

Here $\sigma(K)$ appears naturally in the construction that lifts the projective TQFT to a non-projective one, so no additional invariant needs to be reconstructed from the Berry data\footnote{Invertible $E_8$ stacking data not contained in the projective modular representation are not independently reconstructed.} \cite{Galviz2}. After the Walker--Maslov correction, the natural-isomorphism class of the extended Abelian Chern--Simons TQFT depends only on the finite quadratic module $(G_K,q_K)$ \cite{Galviz4}. Equation~\eqref{eq:intertwine} is therefore the precise sense in which the topological part of metric Berry holonomy can be computed in the real-polarized basis. We do not assert that every microscopic Hamiltonian in an Abelian phase is adiabatically connected to this fixed point.

\vspace{5pt}
\textit{\textbf{Berry tomography of Abelian anyon theories--}}
For the lattice $K$, after geometric quantization of the Abelian Chern--Simons theory, we obtained the discriminant group and quadratic form \cite{Galviz2}
\begin{equation}
G_K=\Lambda^*/K\Lambda,\qquad
q_K([x])=\exp\!\bigl(\pi i\,x^T K^{-1}x\bigr),
\label{eq:qK}
\end{equation}
with nondegenerate symmetric bicharacter
$\Omega_K(a,b)=q_K(a+b)/[q_K(a)q_K(b)]$. In the torus anyon (Bohr--Sommerfeld) basis,
\begin{equation}
S_{ab}=|G_K|^{-1/2}\Omega_K(a,b),\qquad
\Theta_{ab}=\delta_{ab}q_K(a).
\label{eq:ST}
\end{equation}
A physical Dehn twist may contain a common framing factor $T^{\rm phys}=e^{-2\pi i c_-/24}\Theta$; hence the projectively invariant quantities are
\begin{equation}
\chi_a(b)=\frac{S_{ab}}{S_{0b}},\qquad
q(a)=\frac{T^{\rm phys}_{aa}}{T^{\rm phys}_{00}}.
\label{eq:normalized}
\end{equation}
For exact pointed data, the normalized rows $\chi_a$ are characters and obey $\chi_{a\oplus b}=\chi_a\chi_b$.

\textbf{Lemma 1.}
Let $S$ be an $N\times N$ unitary matrix with $S_{0b}=N^{-1/2}$, and set $\chi_a(b)=S_{ab}/S_{0b}$. If the finite set $\{\chi_a\}$ is closed under pointwise multiplication, then it is a finite Abelian subgroup of $U(1)^N$. In particular, closure defines a unique Abelian group law $a\oplus b$ by $\chi_{a\oplus b}=\chi_a\chi_b$.

\textit{Proof.}
Unitarity gives $N^{-1}\sum_b|\chi_a(b)|^2=1$. Closure implies $\chi_a^2=\chi_c$ for some $c$, hence also $N^{-1}\sum_b|\chi_a(b)|^4=1$. With $x_b=|\chi_a(b)|^2$, the two equalities imply $N^{-1}\sum_b(x_b-1)^2=0$, so $|\chi_a(b)|=1$ for every $a,b$. Thus $\{\chi_a\}$ is a finite submonoid of the group $U(1)^N$ containing $\chi_0\equiv1$; every element therefore has finite order and an inverse. Distinct rows are orthogonal, hence the product label is unique. $\square$

\textbf{Theorem 2.}
Let $S$ be an $N\times N$ unitary symmetric matrix with distinguished label $0$. After a common phase normalization suppose $S_{0b}=N^{-1/2}$, and define $\chi_a(b)=S_{ab}/S_{0b}$. Let
$T=\mathrm{diag}(T_a)$ be diagonal unitary with $T_0\neq0$, and set $q_a=T_a/T_0$. Assume: (i) $\{\chi_a\}$ is closed under pointwise multiplication; (ii) for the Abelian group law supplied by Lemma~1, $q_0=1$ and $q_{a^{-1}}=q_a$; and (iii)
\begin{equation}
\frac{q_{a\oplus b}}{q_aq_b}=\chi_a(b)
\label{eq:polarization}
\end{equation}
for all labels. Then $\Omega(a,b):=\chi_a(b)$ is a nondegenerate symmetric bicharacter and $q:G\to U(1)$ is a nondegenerate quadratic form with polarization $\Omega$. Thus $(S,T/T_0)$ are exactly the modular data of the pointed category $\mathcal C(G,q)$ in the convention \eqref{eq:ST}. Conversely every pointed modular datum satisfies (i)--(iii).

Indeed, Lemma~1 supplies the group. Multiplicativity of $\Omega$ in its first variable follows from row multiplication, symmetry of $S$ makes $\Omega$ symmetric, and row orthogonality makes it nondegenerate. Equation~\eqref{eq:polarization}, together with $q(-a)=q(a)$, gives $q(ma)=q(a)^{m^2}$, so $q$ is a quadratic refinement. Once $(G,q)$ is reconstructed, the standard finite Weil representation determines the projective mapping-class-group action \cite{Stirling,GelcaUribe}. The classification of extended Abelian Chern--Simons theories shows that $(G,q)$ determines the complete TQFT up to symmetric monoidal natural isomorphism \cite{Galviz4}. Together with the extended Chern--Simons--Reshetikhin--Turaev equivalence \cite{Galviz3}, this gives \begin{equation} 
(G,q)\quad\Longrightarrow\quad  Z^{CS}_{\mathbb T,K}=Z^{\mathrm{CS}}_{(G,q)} \simeq Z^{\mathrm{RT}}_{\mathcal{C}(G,q)}. \label{eq:TQFTequiv} \end{equation}
Here $\mathbb T=\mathrm t/\Lambda$ is a torus gauge group with $K:\Lambda\times\Lambda\rightarrow\mathbb Z$ an even, integral, nondegenerate pairing. The associated finite quadratic module is as in equation \eqref{eq:qK}.  Furthermore, the reconstructed pointed modular category $\mathcal C(G,q)$ determines, up to equivalence, the associated categorical defect theory of \cite{Galviz6}. With the defect theory and normalizations fixed there, write $\widetilde Z^{\mathrm{CS}}_{\mathcal C(G,q)}.$ The reconstruction therefore takes the form

\begin{equation}
(S,T)
\longrightarrow
(G,q)
\longrightarrow
\mathcal C(G,q)
\longrightarrow
\widetilde Z^{\mathrm{CS}}_{\mathcal C(G,q)},
\end{equation}
with
\begin{equation}
\iota^*\widetilde Z^{\mathrm{CS}}_{\mathcal C(G,q)}
\simeq Z^{\mathrm{CS}}_{(G,q)},
\end{equation}
where $\iota$ includes the undecorated bulk bordisms. Thus torus Berry data determine both the bulk theory and its associated categorical defect structures, including the available condensation sectors, fully gapped boundaries, and self-walls, together with their defect lines and junctions.

For a closed genus-$g$ surface $\Sigma_g$, the Abelian Chern--Simons Hilbert space has a canonical basis indexed by the Bohr--Sommerfeld leaves $\{|a_1,\ldots,a_g\rangle\}_{a_i\in G_K}$, which form a torsor for $G_K^g$, and therefore \begin{equation} \begin{aligned} \mathcal H_{\mathbb T,K}(\Sigma_g) &=\bigoplus_{(a_1,\ldots,a_g)\in G_K^g} \mathbb C\,|a_1,\ldots,a_g\rangle,\\ \dim\mathcal H_{\mathbb T,K}(\Sigma_g)&=|G_K|^g . \end{aligned} \end{equation}

For a bordism $X$ with $\partial X=\Sigma_g$, the extended Chern--Simons theory assigns the state
\begin{equation}
\begin{aligned}
Z^{CS}_{\mathbb T,K}(X)&:=
|\det K|^{m_X}
\frac{\sum_{p\in\operatorname{Tors}H^2(X;\Lambda)}
\sigma_{X,p}\otimes\mu_{X,p}}{\#\operatorname{Tors}H^2(X;\Lambda)},\\
Z^{CS}_{\mathbb T,K}(X)&\in\mathcal H_{\mathbb T,K}(\Sigma_g)\
\end{aligned}
\end{equation}
where
$$
\begin{aligned}
m_X=
\frac14\Bigl(&
\dim H^1(X;\mathbb R)+\dim H^1(X,\partial X;\mathbb R)\\
&-\dim H^0(X;\mathbb R)-\dim H^0(X,\partial X;\mathbb R)
\Bigr)
\end{aligned}
$$
is the normalization exponent. Here $\sigma_{X,p}$ denotes the covariantly constant Chern--Simons section of the prequantum line bundle over the torsion component $\mathcal M_{X,p}(\mathbb T)$, while $\mu_{X,p}$ is the canonical translation-invariant half-density induced by Reidemeister torsion on the corresponding Bohr--Sommerfeld leaf.

For a closed three-manifold $M$, this reduces to the partition function
\begin{equation}
Z^{CS}_{\mathbb T,K}(M)=
|G_K|^{\frac{b_1(M)-1}{2}}
\sum_{p\in {\rm Tors}\,H^2(M;\Lambda)}
\frac{e^{2\pi i\,q(p)}}{|{\rm Tors}\,H^2(M;\Lambda)|}\,,
\end{equation}
up to the  framing normalization, where
$b_1(M)=\dim H^1(M;\mathbb R)$ is the first Betti number. The resulting extended theory depends only on the finite quadratic module $(G,q)$ and is independent of the chosen lattice realization. Thus the Berry data need not reconstruct a unique $K$-matrix: any two lattice realizations of the same finite quadratic module determine naturally isomorphic extended TQFTs.

\vspace{5pt}
\textit{\textbf{Stable Berry tomography under finite measurement errors--}}
Exact algebraic reconstruction is insufficient for realistic applications because Berry matrices obtained from finite-size simulations or experiments always contain errors. Previous numerical studies have shown that extracting modular data from finite systems requires controlling finite-size effects and basis ambiguities \cite{Li2022}. We therefore determine whether the topological order remains recoverable from imperfect torus data. Assume a topological basis has been fixed, for example by Wilson-loop \cite{Witten:1988} or minimally-entangled-state (MES) methods \cite{Zhang2012,Zhu2014}, and let $\wt\chi_a$ denote the normalized measured rows. The error model below begins \emph{after} this normalization; it is therefore a bound on the character rows, not directly on raw entries of a noisy $S$ matrix. Define nearest-row multiplication by
\begin{equation}
a\star b:=\arg\min_c\frac1{\sqrt n}
\|\wt\chi_a\wt\chi_b-\wt\chi_c\|_2.
\label{eq:decoder}
\end{equation}

Because noise destroys exact row closure, we reconstruct each fusion product by nearest-neighbor projection of the pointwise product 
$\tilde \chi_a\tilde \chi_b$ onto the finite set of measured normalized rows; the corresponding max–min residual below measures the worst failure of this set to close under multiplication.

\textbf{Theorem 3.}
Suppose, after a relabeling and projective normalization,
\begin{equation}
\|\wt\chi_a-\chi_a\|_\infty\le\delta
\qquad\text{for every }a\in G.
\label{eq:error}
\end{equation}
Then \eqref{eq:decoder} returns the exact fusion law $a\star b=a\oplus b$ whenever
\begin{equation}
\delta<\delta_*:=\frac{\sqrt{9+2\sqrt2}-3}{2}
\simeq0.2196.
\label{eq:radius}
\end{equation}
The sufficient normalized-row decoding radius is therefore independent of $|G|$, giving a uniform finite-error guarantee for exact fusion recovery.

\textit{Proof.}
Distinct character rows are orthogonal, hence their normalized $\ell^2$ distance is $\sqrt2$. For $c=a\oplus b$, writing the three measured factors as exact unit-modulus factors plus errors of size at most $\delta$ gives
\begin{equation}
\frac1{\sqrt N}\|\wt\chi_a\wt\chi_b-\wt\chi_c\|_2
\le3\delta+\delta^2.
\label{eq:correctbound}
\end{equation}
For $d\neq c$, the triangle inequality gives
\begin{equation}
\frac1{\sqrt N}\|\wt\chi_a\wt\chi_b-\wt\chi_d\|_2
\ge\sqrt2-3\delta-\delta^2.
\label{eq:wrongbound}
\end{equation}
The correct row is uniquely nearest if $3\delta+\delta^2<1/\sqrt2$, which is equivalent to \eqref{eq:radius}. $\square$

This theorem also produces measurable consistency diagnostics
\begin{align}
\Delta_S&=\max_{a,b}\min_c\frac1{\sqrt N}
\|\wt\chi_a\wt\chi_b-\wt\chi_c\|_2,\\
\Delta_T&=\max_{a,b}
\bigl|\wt q(a\star b)-\wt q(a)\wt q(b)\wt\chi_a(b)\bigr|.
\end{align}
If the normalized rows and reduced twists are entrywise $\delta$-close to exact pointed data and $\delta<\delta_*$, then
\begin{equation}
\Delta_S\le3\delta+\delta^2,
\qquad
\Delta_T\le4\delta+3\delta^2+\delta^3.
\label{eq:defects}
\end{equation}
Exact vanishing, together with the hypotheses of Theorem~2, is an intrinsic certificate; small defects quantify departure from the pointed fixed-point structure.

\vspace{5pt}
\textit{\textbf{Examples: reconstruction of a fractional quantum Hall phase--}}
As a concrete physical example, consider the Halperin (221) fractional quantum Hall state
$K=\bigl(\begin{smallmatrix}2&1\\1&2\end{smallmatrix}\bigr)$ and $G_K\cong\Z_3$. With $\omega=e^{2\pi i/3}$, suppose only the torus Berry matrices are available from a microscopic calculation, without prior knowledge of the $K$-matrix.
\begin{equation}
S=\frac1{\sqrt3}
\begin{pmatrix}
1&1&1\\
1&\omega^2&\omega\\
1&\omega&\omega^2
\end{pmatrix},
\qquad
\Theta=\mathrm{diag}(1,\omega,\omega).
\end{equation}
The normalized rows obey $\chi_1^2=\chi_2$ and $\chi_1\chi_2=\chi_0$, immediately reconstructing $\Z_3$ and its quadratic refinement. Hence the same torus data determine the projective Berry representation on every genus-$g$ ground-state space of dimension $3^g$, while Eq.~\eqref{eq:TQFTequiv} determines the associated extended TQFT. Since the defect extension is determined by the same finite quadratic module, these torus data also determine the associated Abelian defect structures.

\medskip

To demonstrate that the reconstruction is not restricted to cyclic
discriminant groups, consider the rank-two Abelian Chern--Simons
theory with $K=
\bigl(\begin{smallmatrix}
2&0\\
0&4
\end{smallmatrix}\bigr)$. The discriminant group is $G_K=
\mathbb{Z}_2\oplus\mathbb{Z}_4$, with eight anyon sectors labelled by
\begin{equation}
a=(a_1,a_2),
\qquad
a_1\in\mathbb{Z}_2,\quad
a_2\in\mathbb{Z}_4 .
\end{equation}
Since $
K^{-1}
=
\bigl(\begin{smallmatrix}
1/2&0\\
0&1/4
\end{smallmatrix}\bigr),$ the quadratic refinement recovered by Theorem~2 from the modular data is

\begin{equation}
q_K(a_1,a_2)
=
\exp\left[
\pi i
\left(
\frac{a_1^2}{2}
+
\frac{a_2^2}{4}
\right)
\right].
\end{equation}

The corresponding bicharacter is
\begin{equation}
\Omega_K(a,b)
=
\frac{q_K(a+b)}
{q_K(a)q_K(b)}
=
\exp\left[
2\pi i
\left(
\frac{a_1b_1}{2}
+
\frac{a_2b_2}{4}
\right)
\right].
\end{equation}
Therefore the modular data take the form
\begin{equation}
S_{a,b}
=
\frac{1}{\sqrt{8}}\Omega_K(a,b),
\qquad
T_{a,b}
=
\delta_{a,b}q_K(a).
\end{equation}

The normalized rows $\chi_a(b)=S_{a,b}/S_{0,b}$, satisfy $\chi_{a+b}=\chi_a\chi_b $, and the reconstruction procedure of Theorem~2 recovers the non-cyclic fusion group $G_K\simeq\mathbb{Z}_2\oplus\mathbb{Z}_4$ together with its quadratic refinement directly from the torus Berry matrices.

\vspace{5pt}
\textbf{Toric code: defects from Torus Berry data}

We now illustrate the defect reconstruction for the toric code \cite{kitaev2003anyons,kitaev2012models}, closely following the functorial defects approach in Abelian Chern-Simons theory \cite{Galviz7}, using only the normalized torus Berry matrices
\begin{equation}
S=\frac12
\begin{pmatrix}
1&1&1&1\\
1&1&-1&-1\\
1&-1&1&-1\\
1&-1&-1&1
\end{pmatrix},
\qquad
\Theta=\operatorname{diag}(1,1,1,-1),
\label{eq:toric_st_data}
\end{equation}
in the ordered basis $(1,e,m,\epsilon)$.  The normalized rows
$\chi_a(b)=2S_{ab}$ are closed under pointwise multiplication:
\begin{equation}
\chi_e^2=\chi_m^2=\chi_\epsilon^2=\chi_1,
\qquad
\chi_e\chi_m=\chi_\epsilon .
\label{eq:toric_row_fusion}
\end{equation}
Theorem~2 therefore reconstructs
\begin{equation}
G\simeq\mathbb Z_2\oplus\mathbb Z_2,
\qquad
\epsilon=e\oplus m,
\label{eq:toric_fusion_group}
\end{equation}
while the reduced twists give
\begin{equation}
q(1)=q(e)=q(m)=1,\qquad q(\epsilon)=-1.
\label{eq:toric_quadratic_form}
\end{equation}
Thus $e$ and $m$ are bosons with mutual phase $-1$, whereas $\epsilon$ is a fermion.  No microscopic lattice realization has been used in this reconstruction. The quadratic module immediately determines the two elementary condensation sectors
\begin{equation}
L_e=\{1,e\},\qquad L_m=\{1,m\}.
\label{eq:toric_lagrangian_subgroups}
\end{equation}
Both are Lagrangian subgroups, so they define the electric and magnetic gapped boundaries.  Their boundary line categories are
\begin{equation}
\mathcal A_e\simeq\mathcal A_m\simeq\operatorname{Vec}_{\mathbb Z_2},
\label{eq:toric_boundary_categories}
\end{equation}
with the two simple boundary lines corresponding to the two cosets of
$L_a$ in $G$.  The bulk-to-boundary functor is
\begin{equation}
F_a(X_x)=B_{[x]},\qquad [x]\in G/L_a,
\label{eq:toric_bulk_boundary_functor}
\end{equation}
and the endpoint multiplicities are
\begin{equation}
\dim\operatorname{Hom}_{\mathcal A_a}
\bigl(F_a(X_x)\otimes B_u,B_v\bigr)
=
\delta_{[x]+u,v}.
\label{eq:toric_endpoint_multiplicity}
\end{equation}
This recovers the usual electric and magnetic boundaries while also specifying their boundary line sectors and junction spaces. The functorial construction gives more information than the list of individual boundaries.  In the folded theory the relevant quadratic module is
\begin{equation}
(G\oplus G,q\oplus q^{-1}).
\label{eq:toric_folded_quadratic_module}
\end{equation}
It has exactly six Lagrangian subgroups:
\begin{equation}
\Gamma_{\mathrm{id}},\qquad
\Gamma_s,\qquad
L_a\times L_b\quad (a,b\in\{e,m\}),
\label{eq:toric_six_walls}
\end{equation}
where $s$ exchanges $e$ and $m$ and fixes $\epsilon$.  The two graph subgroups define invertible walls: the identity wall and the electric--magnetic duality wall.  The four product subgroups define noninvertible walls that factor through the vacuum.  Hence the complete self-wall spectrum contains six elementary walls, rather than only the electric--magnetic twist wall.

This classification has a direct state-space realization. Define
\begin{equation}
w_e=(1,1,0,0),\qquad w_m=(1,0,1,0).
\label{eq:toric_boundary_vectors}
\end{equation}
The four noninvertible walls have transmission matrices
\begin{equation}
W_{ab}=w_a^{\,T}w_b,
\qquad a,b\in\{e,m\},
\label{eq:toric_noninvertible_walls}
\end{equation}
while the two invertible walls are
\begin{equation}
W_{\mathrm{id}}=I_4,\qquad
W_s=
\begin{pmatrix}
1&0&0&0\\
0&0&1&0\\
0&1&0&0\\
0&0&0&1
\end{pmatrix}.
\label{eq:toric_invertible_walls}
\end{equation}
Their composition retains intermediate-channel multiplicities:
\begin{equation}
W_{ab}W_{cd}=|L_b\cap L_c|\,W_{ad},
\qquad
W_s^2=I_4.
\label{eq:toric_wall_composition}
\end{equation}
These multiplicities are invisible if one records only the permutation of
anyon labels, but they are part of the functorial wall composition as in
\cite{LanWangWen2015TC}.

The same framework computes boundary-changing sectors.  If
$\sigma_{ab}$ denotes the unique simple line changing boundary $a$ to
boundary $b$, then
\begin{equation}
\sigma_{ba}\circ\sigma_{ab}
\simeq \mathbf 1_a\oplus t_a,
\qquad a\neq b,
\label{eq:toric_boundary_changing_lines}
\end{equation}
where $\mathbf 1_a$ and $t_a$ are the two simple lines on boundary $a$.
The annulus state-space dimensions are
\begin{equation}
\dim\mathcal H_{\mathrm{TC}}(\mathrm{Ann};a,b)
=
|L_a\cap L_b|
=
\begin{cases}
2,&a=b,\\
1,&a\neq b.
\end{cases}
\label{eq:toric_annulus_dimensions}
\end{equation}
Thus the construction determines not only which boundaries exist, but also
their line categories, boundary-changing lines, annulus degeneracies, and
composition laws.

Finally, the toric code exhibits the distinction between a single bulk
realization and the canonical doubled realization.  Since
\begin{equation}
\mathcal C_{\mathrm{TC}}
\simeq \mathcal Z(\operatorname{Vec}_{\mathbb Z_2}),
\label{eq:toric_center_realization}
\end{equation}
the single toric-code theory is realized as the Turaev--Viro theory of
$\operatorname{Vec}_{\mathbb Z_2}$ \cite[Sec. 6]{Galviz6}.  Using the toric-code modular category
itself as input instead produces
\begin{equation}
\mathcal C_{\mathrm{TC}}\boxtimes
\mathcal C_{\mathrm{TC}}^{\mathrm{rev}},
\qquad
\dim\mathcal H_{\mathrm{dbl}}(\Sigma_g)=16^g,
\label{eq:toric_doubled_realization}
\end{equation}
rather than the single-theory dimension $4^g$.  The full-center
multiplicity matrices are, see \cite[Sec. 10]{Galviz6}, 
\begin{equation}
z^{\mathbf1}=I_4,\qquad
z^{A_e}=w_e^{\,T}w_e,\qquad
z^{A_m}=w_m^{\,T}w_m.
\label{eq:toric_full_centers}
\end{equation}
Therefore the same reconstructed $(G,q)$ controls the bulk theory,
condensation algebras, boundaries, walls, boundary junctions, transmission
amplitudes, and the distinction between single and doubled realizations.
This is the additional content supplied by the functorial defect extension of Ref.~\cite{Galviz6}.

The reconstructed quadratic form also determines the signature class entering the finite quadratic theory through the Gauss--Milgram relation 

\begin{equation} \frac{1}{\sqrt{|G|}}\sum_{a\in G}q(a) = e^{2\pi i\,\sigma(K)/8}.\end{equation} 
Thus the phase of the Walker--Maslov correction is fixed by the reconstructed quadratic module. The integer signature of a particular lattice realization is not itself reconstructed, nor is a unique lattice presentation required. Indeed, different even lattices may realize the same finite quadratic module; by the extended classification theorem they nevertheless determine naturally isomorphic Walker--Maslov-corrected extended Abelian Chern--Simons TQFTs \cite{Galviz4}. The restriction to pointed order is essential: for general non-Abelian categories genus-one modular data need not determine the full category \cite{MignardSchauenburg}. 

In summary, torus Berry matrices reconstruct the finite quadratic module $(G,q)$ and, through the classification of extended Abelian Chern--Simons theories, determine the corresponding all-genus extended TQFT up to symmetric monoidal natural isomorphism. Because the Abelian defect extension is organized by the same quadratic-module data, the reconstruction also determines the associated boundary, domain-wall, and condensation structures. Thus genus-one Berry measurements provide a complete algebraic reconstruction of Abelian topological order, including both bulk and defect sectors. The reconstruction remains exact for normalized-row errors below $\delta_*\equiv0.2196$, providing a stable Abelian realization of the Wen--Keski-Vakkuri completeness principle.

\textit{Acknowledgments--}
The author is grateful to Xiao-Gang Wen for his encouragement to pursue this problem during a visit to Beijing.

\textit{Data availability.--} No data were created or analyzed in this theoretical study.

\bibliography{refs}

\section*{SUPPLEMENT PROOFS} 
\textbf{Proof of Theorem 1.}
We give the geometric argument with the half-form and half-density conventions separated explicitly.

Because $K$ is a nondegenerate real symmetric form, choose a $K$-orthogonal decomposition $\mathfrak t=\mathfrak t_+\oplus\mathfrak t_-$ into positive and negative subspaces and set $\epsilon_K=+1$ on $\mathfrak t_+$ and $-1$ on $\mathfrak t_-$. Then $\epsilon_K^2=1$, $\epsilon_K$ is $K$-self-adjoint, and $g_K(u,v)=K(u,\epsilon_Kv)$ is positive definite. The Hodge inner product
\begin{equation}
\langle\alpha,\beta\rangle_h:=\int_\Sigma\alpha\wedge *_h\beta
\end{equation}
is positive definite on $H^1(\Sigma;\R)$. Since $J_h=j_h\otimes\epsilon_K$, for arbitrary
$v=\sum_i\alpha_i\otimes u_i$ one has
\begin{equation}
\omega_K(v,J_hv)=
(\langle\cdot,\cdot\rangle_h\otimes g_K)(v,v)>0
\qquad(v\neq0),
\label{eq:positive}
\end{equation}
so $J_h$ is compatible with $\omega_K$. This proves positivity on the full tensor product, not only on decomposable vectors.

Let $P_h$ be the invariant K\"ahler polarization associated to $J_h$ and $P_L$ the invariant real polarization associated to $L$. Baier--Mour\~ao--Nunes show that, after choosing compatible half-form data, the BKS pairing maps between invariant nonnegative polarizations of an integral symplectic torus extend to the real boundary, are unitary, and are \emph{transitive}~\cite{Baiermouraonunes}. Denote the resulting half-form map by

\begin{equation}
B^{1/2}_{L,h}:\cH^{\Kah,1/2}_{K,h}\longrightarrow\cH^{\real,1/2}_{K,L}.
\end{equation}
Their construction is symplectically natural; for a mapping class $f$, acting by an integral symplectic automorphism of $H^1(\Sigma;\Lambda)$, one has for compatible metaplectic lifts
\begin{equation}
B^{1/2}_{fL,fh}\,U^{\Kah,1/2}_f
=U^{\real,1/2}_f\,B^{1/2}_{L,h}.
\label{eq:naturality}
\end{equation}

The real-polarized Abelian TQFT is written instead in the half-density convention. In that convention the BKS maps are canonically defined only projectively, with composition defect measured by the Maslov--Kashiwara index~\cite{Galviz2}:
\begin{equation}
\begin{aligned}
F_{L_3L_2}F_{L_2L_1}
&=e^{\pi i\mu_K(L_1,L_2,L_3)/4}F_{L_3L_1},\\
\mu_K&=\sigma(K)\mu_\Sigma.
\end{aligned}
\label{eq:Maslov}
\end{equation}
Manoliu's metaplectic construction is precisely a lift of this projective half-density quantization: after a metaplectic choice, one obtains an ordinary Hilbert space and an honest unitary action, while forgetting that choice recovers the projective half-density space~\cite{Manoliu1}. Consequently, after fixing a metaplectic resolution, the class of $B^{1/2}_{L,h}$ gives a map $B_{L,h}$ between the K\"ahler and half-density real-polarized projective spaces. Its projective class is independent of the phase chosen for the lift, which is all that is required for the statement in $\PU(\cH)$.

Projectivizing Eq.~\eqref{eq:naturality} removes the scalar ambiguity of the lift and gives naturality of $B_{L,h}$. To express the mapping-class operator at a fixed real polarization $L$, compose the geometric action $L\mapsto fL$ with BKS transport back to $L$. In the half-density presentation different compositions differ by the scalar in Eq.~\eqref{eq:Maslov}; in the extended TQFT the $\sigma(K)$ Walker weight contributes the inverse scalar, and in $\PU(\cH)$ the scalar is absent in any case. Thus the fixed-polarization projective operator is exactly the projectivization of the metaplectic operator in Eq.~\eqref{eq:naturality}. The same construction on the K\"ahler side is the BKS holonomy from $h$ to $fh$ followed by the geometric action of $f$. Therefore
\begin{equation}
B_{L,h}\rho^{\Kah}_{K,g}(f)
=\rho^{\real}_{K,g}(f)B_{L,h}
\qquad\text{in }\PU(\cH),
\end{equation}
which proves Theorem~1.  $\square$

No additional metric-dependent projective phase survives. The signature-dependent scalar appearing in a chosen lattice presentation is precisely the Walker--Maslov correction; its relevant phase depends only on $\sigma(K)\bmod 8$ and hence, by Gauss--Milgram, only on the finite quadratic module $(G_K,q_K)$. Thus it does not constitute an additional independently reconstructed invariant. 

\vspace{5pt}

\textbf{Proof of Theorem 2 and the defect bounds.}
Lemma~1 has already established that the normalized rows form a finite Abelian group $G$. Set $\Omega(a,b)=\chi_a(b)$. For fixed $b$, the relation $\chi_{a\oplus c}(b)=\chi_a(b)\chi_c(b)$ makes $\Omega$ multiplicative in its first argument. Since $S$ is symmetric and its vacuum row is constant, $\Omega(a,b)=\Omega(b,a)$, so it is a symmetric bicharacter. If $\Omega(a,b)=1$ for all $b$, then $\chi_a=\chi_0$; orthogonality of distinct rows gives $a=0$, proving nondegeneracy.

Equation~\eqref{eq:polarization} says that the polarization of $q$ is $\Omega$. Taking $b=-a$ and using $q(-a)=q(a)$ gives $\Omega(a,a)=q(a)^2$. Multiplicativity of $\Omega$ then implies
\begin{align}
q((m+1)a)
&=q(ma)q(a)\Omega(ma,a)\\
&=q(a)^{m^2+1+2m}=q(a)^{(m+1)^2},
\end{align}
so $q(ma)=q(a)^{m^2}$ for all integers $m$. Hence $q$ is a quadratic form in the standard metric-group sense, and it is nondegenerate because its polarization is.

Finally assume $|\wt q(a)-q(a)|\le\delta$ in addition to Eq.~\eqref{eq:error}, with all exact factors of unit modulus. For the correctly decoded $c=a\oplus b$,
\begin{equation}
|\wt q(a)\wt q(b)\wt\chi_a(b)-q(a)q(b)\chi_a(b)|
\le3\delta+3\delta^2+\delta^3
\end{equation}
by expanding the product of three perturbed factors. Adding $|\wt q(c)-q(c)|\le\delta$ yields the second bound in Eq.~\eqref{eq:defects}; the first is Eq.~\eqref{eq:correctbound}. $\square$

\end{document}